\documentclass[aps,pra,twocolumn,showpacs,amsmath,amssymb,superscriptaddress,unsortedaddress,floatfix,longbibliography]{revtex4-2}

\usepackage[T1]{fontenc} 
\usepackage[utf8]{inputenc} 

\usepackage{graphicx}
\usepackage{amsmath}
\usepackage{graphics}
\usepackage{amssymb}
\usepackage{layout}
\usepackage{verbatim}
\usepackage{amsfonts,epsfig}
\usepackage[dvipsnames,table]{xcolor} 
\usepackage{hyperref}
\hypersetup{colorlinks, linkcolor={blue!90!black}, citecolor={blue!90!black}, urlcolor={blue!90!black} } 

\usepackage{caption}
\usepackage{subcaption}

\newcommand{\tr}{\textrm{Tr}}

\def\duzomniejsze{<\kern-.7mm<}
\def\duzowieksze{>\kern-.7mm>}

\def\textbf#1{{\bf #1}}
\def\beq{\begin{equation}}
\def\eeq{\end{equation}}
\def\be{\begin{equation}}
\def\ee{\end{equation}}
\def\ben{\begin{eqnarray}}
\def\een{\end{eqnarray}}
\def\beqa{\begin{eqnarray}}
\def\eeqa{\end{eqnarray}}
\def\eea{\end{array}}
\def\bea{\begin{array}}
\def\tr{\mathrm{Tr}}

\begin{document}
	
	\title{A signature of quantumness in pure decoherence control}

\author{Bartosz Rzepkowski}
\affiliation{Department of Theoretical Physics, Wroc{\l}aw University of Science and Technology,
	50-370 Wroc{\l}aw, Poland}
	
	\author{Katarzyna Roszak}
	\affiliation{Institute of Physics (FZU), Czech Academy of Sciences, Na Slovance 2, 182 21 Prague, Czech Republic}
	
	\date{\today}

	\begin{abstract}
		We study a decoherence reduction scheme that involves an intermediate measurement on
		the qubit
		in an equal superposition basis, in the general framework of all qubit-environment
		interactions that lead to qubit pure decoherence. We show under what circumstances
		the scheme always leads to a gain of coherence on average, regardless of the time at 
		which the measurement is performed, demonstrating its wide range of applicability. Furthermore, we find that observing an average 
		loss of coherence is a highly quantum effect, resulting from non-commutation
		of different terms in the Hamiltonian. We show the diversity of
		behavior of coherence as effected by the application of the scheme, which is skewed
		towards gain rather than loss, on a variant of the spin-boson model that does
		not fulfill the commutation condition.  
	\end{abstract}

	\maketitle

\section{Introduction \label{sec:introd}}

The results of Ref.~\cite{roszak15b} show that if an excitonic qubit confined in a quantum dot
interacting with a bath of phonons undergoes a procedure where the qubit is initialized in a superposition
state, decoheres for a time which is longer than the few-picosecond time-scale of the phonon-induced
decoherence \cite{borri01,krummheuer02,vagov04,vagov06,mermillod16,wigger18}, and then is measured in the equal-superposition basis of qubit pointer states,
the post-measurement decoherence is, on average, smaller than the standard decoherence 
(the decoherence one would observe if no measurement was performed).
The fact that the decoherence observed is different is reasonably easy to understand.
The exciton-phonon interaction leads to entanglement being formed for all thermal-equilibrium 
phonon states at finite temperatures \cite{roszak15,rzepkowski21}. Since for pure-decoherence
the generation of entanglement is equivalent to the generation of quantum discord \cite{roszak17},
a measurement on one subsystem will result in a discernible change of the state of the other subsystem 
and this change qualitatively depends on the measurement outcome 
\cite{ollivier01,modi12,modi14}.
Why the effect should on average be advantageous for retaining qubit coherence
is not so obvious and we will explore it here in the general framework of all
interactions that lead to qubit pure decoherence due to an interaction with some
environment. 

In the following we will answer the questions: What are the conditions on the system for
the procedure to counter the decoherence on average?
Is qubit-environment entanglement necessary? What does average coherence loss
signify?

The answers to these questions are important both from a utilitarian point of view as well as
from a purely theoretical standpoint. On one hand, testing a simple procedure which entails only
straightforward operations and measurements performed solely on the qubit which can reduce decoherence
is important, especially understanding the conditions of its applicability. From the theory side, it is important
to understand where the quantum nature of the environment is crucial, because these are the
situations when the classical description of noise will yield wrong results. 

We study pure decoherence, because this is the broadest level of considerations which allows to draw
definite conclusions, while still encompassing a large number of qualitatively different
environments. 
This includes decoherence which is the result of entanglement 
\cite{salamon17,rzepkowski21,strzalka20} as well as 
sources of noise which do not require the establishment of quantum correlations \cite{pernice11,lofranco12}, Markovian and non-Markovian \cite{lofranco12,szankowski17,chen18,wenderoth21,chruscinski22}
processes, as well as pure and mixed initial environmental states. 
It is also the dominating decohrence mechanism for many 
state of the art solid state qubit realizations
\cite{cywinski09,paladino14,wu18,touzard19,campagne20,wood18,tchebotareva19,wang20b,watzinger18,hendrickx20,miao20,madzik21,fricke21}
and as such, methods for pure decoherence control are of contemporary relevance.

We find that qubit-environment entanglement is in fact not necessary for the operation
of the scheme. It relies on the memory of the environment which is affected by the 
joint evolution with the qubit and on the transfer of information about the qubit
into the environment which is the outcome of the qubit being measured. Similarly as in the
case of the spin echo \cite{roszak21}, the properties of the operators which
determine the evolution of the environment in the presence of the pointer states of 
the qubit are critical. If these operators commute, application of the decoherence
control procedure will never yield loss of coherence. Similarly, if the operators 
commute with the initial state of the environment (but not necessarily with each other),
then the coherence will be increased or remain the same
(depending on the time when the measurement is applied).

We test our findings on the same system as in ref.~\cite{roszak15b} which is an
asymmetric variant of solid-state
realizations of the spin-boson model, but we find that if the interval between initialization and the
measurement is small, the average outcome of the procedure can be negative. This is because
the unitary operators responsible for the evolution of the environment do not commute. 
At large delay times only an increase in mean coherence
is possible because the environment is sufficiently large that the bosonic
creation and annihilation operators behave as if they would commute and the non-classical phases that are the
effect of the lack of commutation, cancel out. The effect is in agreement
with the notion that with growing system size one should expect more classical behavior. 
This behavior is not observed for small environments, as we demonstrate
using an analogous qubit-environment system with a discrete spectrum of few bosonic modes.

Average loss of coherence is therefore a signature of observable quantum behavior
of the environment under the influence of the qubit. Its occurrence is only possible when
specific terms in the Hamiltonian, which are observables on the environment, do not 
commute. It corresponds to the same situations when entanglement with the environment
can be observed by operations and measurements on the qubit \cite{roszak19a,rzepkowski21,strzalka21},
when the spin echo is not a good way of countering decoherence \cite{roszak21},
and when different measurement outcomes lead to a different degree of coherence
of the teleported state
in repeated noisy teleportation \cite{harlender22}, e.~g.~ in a quantum network scenario
\cite{roszak22}.

The paper is organized as follows. In Sec.~\ref{sec:scheme}
we specify the class of Hamiltonians under study and describe the decoherence
reduction scheme in detail. In Sec.~\ref{sec3} we present our findings for this class
of Hamiltonians, including the upper and lower bounds on average coherence gain
and the study of conditions that guarantee that the gain is positive. 
Sec.~\ref{sec4} contains results for a system that can support coherence loss
with emphasis
on effects connected with the size of the environment. Sec.~\ref{sec5} concludes the 
article.

	\section{Scheme for decoherence reduction} \label{sec:scheme}
	
	\subsection{Pure decoherence}
	
	We are investigating a general class of qubit-environment Hamiltonians that lead to pure decoherence \cite{roszak18,chen18,chen19,popovic21}. 
	Such Hamiltonians can always be written in the form
	\begin{equation} \label{ham}
	    \hat{H} = \sum_{i=0,1} \varepsilon_i |i\rangle\langle i | + \hat{H}_E + \sum_{i=0,1} |i\rangle\langle i | \otimes \hat{V}_i,    
	\end{equation}		
	where the first term describes the free evolution of the qubit, with $\varepsilon_i$ denoting the energies of the qubit pointer states $|0\rangle$ and $|1\rangle$, $\hat{H}_E$ is the free Hamiltonian of the environment, and the
	last term describes their interaction. The environmental operators $\hat{V}_i$ are responsible for the effect that the qubit in pointer state $|i\rangle$
	has on the environment. This conditional effect of the qubit on the environment is the source of
	pure decoherence, which for pure states is interpreted as the result of information about the state of the
	qubit leaking into the environment \cite{zurek03}.
	
	Any Hamiltonian of the form (\ref{ham}) yields a qubit-environment evolution operator which can be written
	in the form ($\hbar = 1$)
	\begin{equation}
	\label{u}
	    \hat{U}(t) = \sum_{i=0,1} e^{-i\varepsilon_i t }|i\rangle\langle i | \otimes \hat{w}_i(t),
	\end{equation}
	where the conditional operators acting on the environment are given by 
	\begin{equation}
	    \label{w}
	    \hat{w}_i(t) =  e^{-i \left(\hat{H}_E + \hat{V}_i\right)t }.
	\end{equation}
	
	\subsection{The Scheme}
	
	The protocol described in Ref.~\cite{roszak15b}, used to decrease phonon-induced decoherence
	of exciton qubits by decoherence itself, involves first preparing the environment by a controlled decoherence 
	process on the qubit before the actual undesirable decoherence process takes place. To this end, the qubit
	is prepared in an equal superposition of its pointer states (to maximize the effect that the qubit
	has on the environment),
	$|+\rangle = \frac{1}{\sqrt{2}}(|0\rangle + |1\rangle)$, and then it evolves in the presence of the 
	environment for time $\tau$. At time $\tau$ the qubit is measured in the 
	equal-superposition basis, $|\pm\rangle$, yielding a product of one of the two qubit states and 
	a corresponding new state of the environment. The qubit can now be transformed
	to the desired initial state and the post-measurement decoherence, which is still governed by the same exciton-phonon Hamiltonian, is
	affected by the new ``initial'' state of the environment. 
	In Ref.~\cite{roszak15b} it has been shown that on average (over the measurement outcomes), the pure decoherence
	observed post-measurement is smaller or equal to the decoherence which is usually obtained
	for an environment initially at thermal equilibrium if the preparation time $\tau$ is long enough
	to allow a steady state to be reached. 
	
	We will be studying exactly the same protocol, but in a general pure decoherence scenario
	in order to understand the range of applicability of the scheme of Ref.~\cite{roszak15b},
	as well as the origins of the effect (why negative effects were not observed). 
	
	The preparation of the environment starts with the enviornment in state
	$\hat{R}(0)$ and qubit in state $|+\rangle$, so at time $\tau$ the qubit-environment density matrix
	obtained using the evolution operators (\ref{u})
	is given by 
    \begin{equation}
        \label{sigma}
    	\hat{\sigma}(\tau) = \frac{1}{2} \left(\begin{matrix} \hat{R}_{00}(\tau) & e^{-i\Delta\varepsilon \tau }\hat{R}_{01}(\tau) \\ 
    	e^{i\Delta\varepsilon \tau }\hat{R}_{10}(\tau) & \hat{R}_{11}(\tau) \end{matrix}
    	\right),
	\end{equation}
	where $\Delta\varepsilon=\varepsilon_0-\varepsilon_1$ and the matrices responsible for the degrees of freedom of the environment are given by
\begin{equation}
    	\label{rij}
    	\hat{R}_{ij}(\tau) = \hat{w}_i(\tau) \hat{R}(0) \hat{w}_j^\dagger(\tau),
	\end{equation}	
	$i,j = 0,1$.
	In eq.~(\ref{sigma}) the matrix form pertains to the qubit subsystem, while the environmental degrees
	of freedom are taken into account by the $\hat{R}_{ij}(\tau)$ matrices. This notation is particularly
	convenient for qubit-environment systems undergoing pure decoherence,
	since the qubit is the system of interest and there is a large asymmetry in its size as compared 
	to the environment, which is of arbitrary dimension.
	
	For one, it makes tracing out environmental degrees of freedom
	to obtain the state of the qubit particularly straightforward, since the partial
	trace conserves the matrix form of eq.~(\ref{sigma}). Hence, the density matrix of the qubit
	$\hat{\rho}(\tau)=\tr_{\mathrm{E}}\hat{\sigma}(\tau)$ does not display any evolution of its diagonal
	elements, since the matrices $\hat{R}_{00}(\tau)$ and $\hat{R}_{11}(\tau)$ are density matrices
	(with unit trace), while the off-diagonal elements evolve according to 
	\begin{equation}
	    \label{rho01}
    	\hat{\rho}_{01}(\tau)=\langle 0|\hat{\rho}(\tau)|1\rangle = \frac{1}{2}e^{-i\Delta\varepsilon \tau } \tr\hat{R}_{01}(\tau).
	\end{equation}
	
	If the qubit did undergo decoherence due to the interaction with the environment, so $|\hat{\rho}_{01}(\tau)|\neq 1$,
	then the qubit-environment density matrix (\ref{sigma}) cannot be written in product form and some correlations
	between the two subsystems must have been formed. For pure initial states, these correlations must be 
	entanglement \cite{zurek03}, but for mixed states both classical and quantum correlations can be
	the source of decoherence, depending on the nature of the interaction (\ref{ham}) and the initial state of 
	the environment, $\hat{R}(0)$. The if and only if condition of separability is given by \cite{roszak15}
	\begin{equation}
	\label{sep}
	\left[\hat{R}(0),\hat{w}_0^\dagger(\tau)\hat{w}_1(\tau)\right]=0.
	\end{equation}
	
	Regardless of the nature of the decoherence, the effect of the measurement on the qubit in the
	$|\pm\rangle=\frac{1}{\sqrt{2}}\left(|0\rangle\pm|1\rangle\right)$ basis
	is described in the same way, and the post measurement qubit-environment state is given by
	\begin{equation}
	\label{sigmapm}
	\sigma_{\pm}(\tau)=|\pm\rangle\langle\pm |\otimes\hat{R}_{\pm}(\tau),
	\end{equation}
	where the index $\pm$ distinguishes between the two measurement outcomes.
	The unnormalized post-measurement state of the environment is given by
	\begin{eqnarray}
	\label{run}
    \tilde{R}_{\pm}(\tau) &=& \frac{1}{4}\left[\hat{R}_{00}(\tau) +\hat{R}_{11}(\tau)\right.\\
    \nonumber
    &&\left.{\pm} \left(e^{-i\Delta\varepsilon \tau }\hat{R}_{01}(\tau) + e^{i\Delta\varepsilon \tau }\hat{R}_{10}(\tau) \right)\right].
    \end{eqnarray}
    The probabilities of obtaining each measurement outcome are
    \begin{equation}
    \label{prob}
    p_{\pm}(\tau)=\tr\tilde{R}_{\pm}(\tau),
    \end{equation}
    and the normalized post-measurement density matrices of the environment 
    are given by
    $
    \hat{R}_{\pm}(\tau)=\tilde{R}_{\pm}(\tau)/p_{\pm}(\tau)$.
    
    It is now relevant to note that although the post-measurement state is of product form,
    so it does not contain any correlations, neither quantum nor classical,
    the new state of the environment now contains information about the pre-measurement state of the qubit. 
    This manifests itself by the phase factors in the second line of eq.~(\ref{run})
    which are the outcome of the free evolution of the qubit.
    Incidentally, if the initial qubit state was not an equal superposition state, this would also be
    visible in the post-measurement state of the environment, which would contain a different mixture of the four 
    components in the matrix.
    
    Now that the environment has been prepared, it is time to prepare the initial state of the qubit,
    $|\psi\rangle = \alpha|0\rangle +\beta |1\rangle$, yielding the qubit-environment state
    \begin{equation}
    \label{sigmapm2}
    \sigma'_{\pm}(\tau)=|\psi\rangle\langle\psi |\otimes\hat{R}_{\pm}(\tau),
    \end{equation}
    depending on the measurement outcome. 
    
    The post-measurement evolution is governed by the same Hamiltonian
    (\ref{ham}), but as a function of the post-measurement time $t$ and with
    a new initial state given by eq.~(\ref{sigmapm2}).
    Hence, the $t$ evolution mirrors the evolution in eq.~(\ref{sigma}), but in the now $t$-dependent 
    environmental matrices (\ref{rij}), the initial state of the environment, $\hat{R}(0)$, is now replaced by 
    $\hat{R}_{+}(\tau)$ for measurement outcome $|+\rangle$ and by $\hat{R}_{-}(\tau)$
    for measurement outcome $|-\rangle$, and the qubit state can be any superposition of pointer states.
    The full system density matrix at time $t$ is given by
    \begin{equation}
    \label{sigmat}
    \hat{\sigma}_{\pm}(\tau,t) = \left(\begin{matrix} |\alpha|^2\hat{R}_{00}^{\pm}(\tau,t) &  \alpha\beta^*e^{-i\Delta\varepsilon t }\hat{R}_{01}^{\pm}(\tau,t) \\ 
    \alpha^*\beta e^{i\Delta\varepsilon t }\hat{R}_{10}^{\pm}(\tau,t) & |\beta|^2 \hat{R}_{11}^{\pm}(\tau,t) \end{matrix}
    \right),
    \end{equation}
    where 
    \begin{equation}
    \label{rijt}
    \hat{R}_{ij}^{\pm}(\tau,t) = \hat{w}_i(t) \hat{R}_{\pm}(\tau) \hat{w}_j^\dagger(t).
    \end{equation}
    
    As before, the qubit decoherence resulting from the joint QE evolution given by eq.~(\ref{sigmat})
    is pure dephasing, so tracing out of the environmental degrees of freedom
    leaves the qubit occupations constant, while the coherences evolve according to
    \begin{equation}
    \label{rhot}
    \rho_{\pm}^{01}(\tau,t)= 
    \alpha\beta^* e^{-i\Delta\varepsilon t } \tr\hat{R}_{01}^{\pm}(\tau,t).
    \end{equation}
    
    \section{Gain of coherence \label{sec3}}
    
    To quantify how the application of the scheme can help preserve qubit coherence we will be studying
    the degree of coherence defined as the absolute value of the off-diagonal element of the density matrix of 
    the qubit divided by its initial value
    \begin{equation}
    \label{degree_of_coherence}
    D_{\pm}(\tau,t) = |\rho_{\pm}^{01}(\tau,t)| / |\alpha\beta|=\left|\tr\hat{R}_{01}^{\pm}(\tau,t)\right|.
    \end{equation}
    
    This quantity differs depending on the measurement outcome and may also be used to quantify the 
    amount of coherence present in the qubit when no decoherence-reduction scheme was applied
    (standard decoherence)
    by simply setting the initial decoherence time $\tau$ to zero,
    \begin{equation}
    \label{dst}
    D(t)=D_{+}(0,t)=D_{-}(0,t).
    \end{equation}
    It is often convenient to study the degree of coherence relative to the amount of coherence that would be present
    at time $t$ with the environmental state given by $\hat{R}(0)$,
    \begin{equation}
    \label{relative_gain}
    g_{\pm}(\tau,t) = D_{\pm}(\tau,t) - D(t).
    \end{equation}
    We will call this quantity the coherence gain following Ref.~\cite{roszak15b}. Positive values of the gain
    mean that coherence at time $t$ is grater than for standard decoherence,
    but situations when this quantity is negative are also possible (and common). 
    
    As the most interesting effect found in Ref.~\cite{roszak15b} and the one we wish to explore is the fact
    that asymptotic gains of coherence for the exciton-phonon system are non-negative on average,
    we will also be using coherence gain averaged over the two measurement outcomes,    
    \begin{equation}
    \label{average_gain}
    g_{av}(\tau,t) = p_+(\tau) g_+(\tau,t) + p_-(\tau) g_-(\tau,t).
    \end{equation}
    
    Using eqs (\ref{run}), (\ref{prob}), and (\ref{rijt}) it is straightforward to find the explicit formulas 
    for the above quantities. Especially relevant are
    \begin{equation}
    \label{czesci}
    p_{\pm}(\tau) D_{\pm}(\tau,t)=\left|A(\tau,t)\pm B(\tau,t)\right|,
    \end{equation}
    where    
    \begin{subequations}
    	\label{czesciab}
    \begin{eqnarray}
    \label{czescia}
    A(\tau,t)&=&\frac{1}{4} \tr\left[\hat{w}_0(t)\left(\hat{R}_{00}(\tau) + \hat{R}_{11}(\tau)
    \right)\hat{w}_1^\dagger(t))\right],\\
    \nonumber
    B(\tau,t)&=&\frac{1}{4} \tr\left[\hat{w}_0(t)\left(
    e^{-i\Delta\varepsilon\tau}\hat{R}_{01}(\tau) + e^{i\Delta\varepsilon\tau}\hat{R}_{10}(\tau) \right)\hat{w}_1^\dagger(t))\right].\\
    \label{czescib}
    \end{eqnarray}	
\end{subequations}	
	This is because one can write the average gain of coherence as
	\begin{equation}
	\label{eq:average_gain}
	g_{av}(\tau,t) = D_{av}(\tau,t)-D(t),
	\end{equation}
	where the average degree of coherence is given by
	\begin{equation}
	\label{dav}
	D_{av}(\tau,t)=p_+(\tau) D_+(\tau,t) + p_-(\tau) D_-(\tau,t).
	\end{equation}
	This form, together with eqs (\ref{czesci}) and (\ref{czescia}), allows to analyze the possibility for $g_{av}(\tau,t)$ to have
	negative values on a more general level, due to the evident symmetry between the 
	terms in eq.~(\ref{dav}), which results
	in the following inequality 
	\begin{equation}
	\label{davin}
	D_{av}(\tau,t)\ge 2\max\{|A(\tau,t)|,|B(\tau,t)|\}.
	\end{equation}
	This means that the study of the occurrence of negative values
	in the evolution of $g_{av}(\tau,t)$ can be reduced to
	the comparison of the greater of the functions
	$|A(\tau,t)|$ and $|B(\tau,t)|$ and the 
	degree of coherence in the standard evolution, $D(t)$.
	
	Equality in (\ref{davin}) is realized when the phases of
	$A(\tau,t)$ and $B(\tau,t)$ align, meaning that 
	$e^{i\phi_A(\tau,t)}=\pm e^{i\phi_B(\tau,t)}$ in
	\begin{subequations}
		\label{fazy}
		\begin{eqnarray}
		\label{fazya}
		A(\tau,t)&=&e^{i\phi_A(\tau,t)}|A(\tau,t)|,\\
		\label{fazyb}
		B(\tau,t)&=&e^{i\phi_B(\tau,t)}|B(\tau,t)|.
		\end{eqnarray}
	\end{subequations}
	
	Incidentally the maximum average degree of coherence is limited by
	\begin{equation}
	\label{davin2}
	D_{av}(\tau,t)\le 2\sqrt{|A(\tau,t)|^2+|B(\tau,t)|^2}.
	\end{equation}
	Equality is obtained when $e^{i\phi_A(\tau,t)}=\pm i e^{i\phi_B(\tau,t)}$.
	
	\subsection{Fast free qubit evolution}
	
	A reasonable assumption in the study of qubit decoherence is that the energy difference
	between the qubit states is much larger than the energy scales responsible for the interaction
	with the environment, which means that the free evolution of the qubit is much faster 
	than any evolution resulting from the decoherence process. This means that the phase 
	factors $e^{\pm i\Delta\varepsilon\tau}$ in eq.~(\ref{czescib}) drive the $\tau$-dependence
	in the average gain and degree of coherence, while the rest of the $\tau$-dependence
	and $t$-dependence in eqs (\ref{czescia}), (\ref{czescib}), and (\ref{dst})
	are comparatively slow. In this situation, one can assume that the fast evolution
	yields all possible values of the phase factor while 
	all other factors remain constant. This means that the minimum and maximum values of 
	the average quantities corresponding to a given preparation time $\tau$ and decoherence 
	time $t$ will be realized and one can limit the study to the envelope functions of 
	the average gain and degree of coherence. 
	
	The conditions for the envelope functions of the gain of coherence 
	can be written as 
	\begin{subequations}
		\label{cond12}
		\begin{eqnarray}
		\label{cond1}
		\sin\left(\phi_B(\tau,t)-\phi_A(\tau,t)\right)&=&0,\\
		\label{cond2}
		\cos\left(\phi_B(\tau,t)-\phi_A(\tau,t)\right)&=&0,
		\end{eqnarray}
	\end{subequations}
	where the first guarantees equality in eq.~(\ref{davin}) and
	the second in eq.~(\ref{davin2}).
	As the quantity $A(\tau,t)$ has no dependence on the free qubit evolution, we will be taking
	into account only the evolution of
	$B(\tau,t)$, which is explicitly given in eq.~(\ref{czescib}), to find the fast
	$\tau$-dependence of
	$\phi_B(\tau,t)$. 
	
	We define the components in eq.~(\ref{czescib}) as
	\begin{equation}
	B_{ij}(\tau,t)=
	\frac{1}{4} \tr\left[\hat{w}_0(t)
	\hat{R}_{ij}(\tau)\hat{w}_1^\dagger(t))\right]
	\end{equation}
	and their internal phases $\exp[i\phi_{ij}(\tau,t)]=B_{ij}(\tau,t)/|B_{ij}(\tau,t)|$.
	Using this notation, we get conditions equivalent to (\ref{cond12})
	in the form
	\begin{subequations}
		\label{cond1020}
		\begin{eqnarray}
		\label{cond10}
		B_{+}
		\sin\left[\left(\phi_{+}\right)-\phi_A\right]
		\cos\left[\left(\phi_{-}\right)-\Delta\varepsilon\tau\right]&&\\
		\nonumber
		+B_{-}
		\sin\left[\left(\phi_{-}\right)-\Delta\varepsilon\tau\right]&=&0,\\
		\label{cond20}
		B_{+}
		\cos\left[\left(\phi_{+}\right)-\phi_A\right]
		\cos\left[\left(\phi_{-}\right)-\Delta\varepsilon\tau\right]&=&0.
		\end{eqnarray}
	\end{subequations}
	Here, we have suppressed the explicit time dependence in all factors
	except for the $\Delta\varepsilon \tau$ dependence for conciseness.
	We also denoted the sum and difference of the $B_{ij}$ terms,
	$B_{\pm}=B_{10}\pm B_{01}$,
	and of the intrinsic phases in $B_{ij}$, $\phi_{\pm}=\phi_{10}\pm \phi_{01}$.
	
	Although the conditions (\ref{cond1020}) are always applicable, they are primarily useful
	in the situation under study. Eq.~(\ref{cond10}) signifies when the lower bound
	of the average degree of coherence (and hence, coherence gain) is reached. Even though it has two
	terms, it is obvious that the function on the left hand side must cross zero during
	its evolution with $\Delta\varepsilon\tau$. 
	The same is true for eq.~(\ref{cond20}), which is responsible for the average degree and gain of coherence reaching its upper bound. Hence, for all times $\tau$ and $t$ there
	exist phase values, which guarantee equality in inequalities (\ref{davin}) and (\ref{davin2}),
	so the functions on their right sides are in fact the envelope functions 
	and not just arbitrary bounds. 
	
	\subsection{Commutation relations that guarantee coherence gain}
	
	For pure decoherence, a special class of Hamiltonians is distinguished by the fact that the conditional evolution
	operators of the environment (\ref{w}) commute at all times,	
	\begin{equation}
	\label{comut}
	\left[\hat{w}_0(t), \hat{w}_{1}(t')\right] = 0.
	\end{equation}
	The simplest situation when eq.~(\ref{comut}) is fulfilled is when all environmental terms in the full 
	Hamiltonian (\ref{ham}) commute with each other, meaning that 
	\begin{equation}
	\left[\hat{V}_0,\hat{V}_1\right]=\left[\hat{V}_0,\hat{H}_E\right]=\left[\hat{V}_1,\hat{H}_E\right]=0,
	\end{equation}	
	but this is not the only one. 
	
	Note that commuting environmental operators are a sign of certain type of classicality,
	as there is no equivalent of non-commutation for observables in classical physics. 
	Hence, such environments lead to decoherence that will behave differently than
	in the situation when the commutation criterion (\ref{comut}) is not fulfilled. 
	For example, even though in this case decoherence can be
	accompanied by the generation of qubit-environment entanglement \cite{roszak15},
	schemes for the detection of this type of entanglement by operations and measurements
	on the qubit alone will not work \cite{roszak19a,rzepkowski21}. 
	Entanglement means that decoherence is accompanied by the transfer of information
	about the qubit state into the environment \cite{roszak19b}, but for commuting environmental operators
	this information would have to be read out directly from the state of the environment,
	as it does not manifest itself due to back-action in the evolution of the qubit. 
	
	For the scheme under study, commutation (\ref{comut}) guarantees that for all times
	$t$ and $\tau$, the quantity defined in eq.~(\ref{czescia}) 
	has no $\tau$ dependence and is proportial to the standard decoherence,
	\begin{equation}
	\label{acomut}
	A(\tau,t)=\frac{1}{2}\tr\left[\hat{w}_0(t)\hat{R}(0)\hat{w}_1^{\dagger}(t)\right].
	\end{equation}
	This means that the average degree of coherence is alwas greater or equal to 
	the degree at time $t$ when the scheme was not applied, 
	$D_{av}(\tau,t)\ge D(t)$
	and consequently the gain in coherence is always non-negative 
	(coherence on average is always increased by the application of the scheme),
	$g_{av}(\tau,t)\ge 0$.

	This situation is optimal when using the scheme to counter decoherence, as in the 
	worst-case scenario, no effect will be obtained, but as we will show in the example 
	in the following section, the fast oscillations of the $\tau$-dependent phase factor 
	in the quantity of eq.~(\ref{czescib}) lead to fast oscillations in the average coherence
	gain as a function of preparation time $\tau$, so that most values of $\tau$ yield an actual
	gain. 
	
	Note that for commuting environmental evolution operators, the other term which enters
	the average degree of coherence (\ref{czescib}) also takes a simpler form,
	\begin{eqnarray}
	\label{bcomut}
	B(\tau,t)&=&\frac{1}{4} \{e^{-i\Delta\varepsilon\tau}
	\tr\left[\hat{w}_0(t+\tau)\hat{R}(0)\hat{w}_1^\dagger(t+\tau))\right]\\
	\nonumber
	&&+e^{i\Delta\varepsilon\tau}
	\tr\left[\hat{w}_0(t-\tau)\hat{R}(0)\hat{w}_1^\dagger(t-\tau))\right]\}.
	\end{eqnarray}
	Now, the term is constructed from terms which are easily interpreted as decoherence factors
	corresponding to a later time and an earlier time, shifted by the preparation time $\tau$.
	
	It is relevant to note here that the commutation of environmental evolution operators
	is not the only situation when $g_{av}(\tau,t)\ge 0$ for all times $\tau$ and $t$
	and the operation of the scheme for decoherence control works optimally. Another situation is
	when both operators commute with the initial state of the environment
	\begin{equation}
	\label{comut2}
	\left[\hat{w}_0(t),\hat{R}(0)\right]=\left[ \hat{w}_{1}(t),\hat{R}(0)\right] = 0,
	\end{equation}
	so that 
	\begin{equation}
	\hat{R}_{00}(\tau)=\hat{R}_{11}(\tau)=\hat{R}(0).
	\end{equation}
	This does not necessarily require the commutation of the evolution operators (\ref{comut})
	and may be the consequence of the form of the state of the environment,
	as in the case of an infinite-temperature density matrix which commutes with any
	operators. Incidentally, contrarily to the situation when commutation criterion (\ref{comut})
	is fulfilled, criterion (\ref{comut2}) being met directly implies that there is no 
	entanglement generated between the qubit and the environment during the initial 
	evolution till time $\tau$.

    \section{Charge qubit and phonon environment \label{sec4}}

    In the following section, we will study the average gain of coherence
    with special attention given to situations when negative values can be observed.
    We follow Ref.~\cite{roszak15b} and study 
    an excitonic quantum dot qubit interacting with phonons,     
    but we will take into account two variants:
    Firstly we consider a bath of bulk phonons as in Ref.~\cite{roszak15b},
    but we show that for small times negative gains are observed even though
    there is a continuous spectrum of phonons. Secondly we restrict the number
    of phonon modes to a relatively small number, to study finite environment effects. 
    
    We choose this example of a qubit-environment system precisely because of its versatility.
    Although the exciton-phonon Hamiltonian is closely related to the spin-boson model 
    \cite{tuziemski19,lambert19,wenderoth21,miessen21,chruscinski22},
    its environmental evolution operators $\hat{w}_i(t)$ do not commute, yielding an additional
    non-classical phase factor on top of decoherence concurring with the spin-boson 
    model decoherence. For bulk phonons initially at thermal equilibrium, this phase factor
    averages out to zero (on faster time scales than the decoherence takes place),
    but if only a limited number of phonon modes is taken into account, the cancellation
    does not take place. This results in a multitude of different scenarios that can be 
    taken into account, which lead to comparable results in case of standard decoherence,
    but have a qualitative impact on the effects observed in the decoherence control
    protocol under study. 
    
    For the excitonic qubit \cite{borri01,vagov04,michaelis10}
    the $|0\rangle$ state means that the dot is empty, while the $|1\rangle$ state means
    that there is an exciton in its ground state in the quantum dot. Hence, phonons couple only to
    the $|1\rangle$ qubit state and the interaction is naturally asymmetric. 
    The Hamiltonian of the
    system is given by eq.~(\ref{ham}) with 
    the eigenenergies correponding to the pointer states
    $\varepsilon_0=0$ and $\varepsilon_1=\varepsilon$, where $\varepsilon$ is the excitonic
    ground state energy. The free Hamiltonian of the environment is given by
    $\hat{H}_E=\sum_{\pmb{k}} \omega_{\pmb{k}} \hat{b}_{\pmb{k}}^\dagger \hat{b}_{\pmb{k}}$, where
    $\hat{b}_{\pmb{k}}^\dagger$ and $\hat{b}_{\pmb{k}}$ are phonon creation and annihilation operators in mode $\pmb{k}$ and $\omega_{\pmb{k}}$ are the corresponding energies. 
    The environmental operators in the interaction term are given by $\hat{V}_0=0$
    (phonons do not interact with the exciton when it's not there) and 
    $\hat{V}_1=\sum_{\pmb{k}} (f_{\pmb{k}}^*\hat{b}_{\pmb{k}} + f_{\pmb{k}} \hat{b}_{\pmb{k}}^\dagger )$.
    
\begin{figure}
	\includegraphics[width=\linewidth]{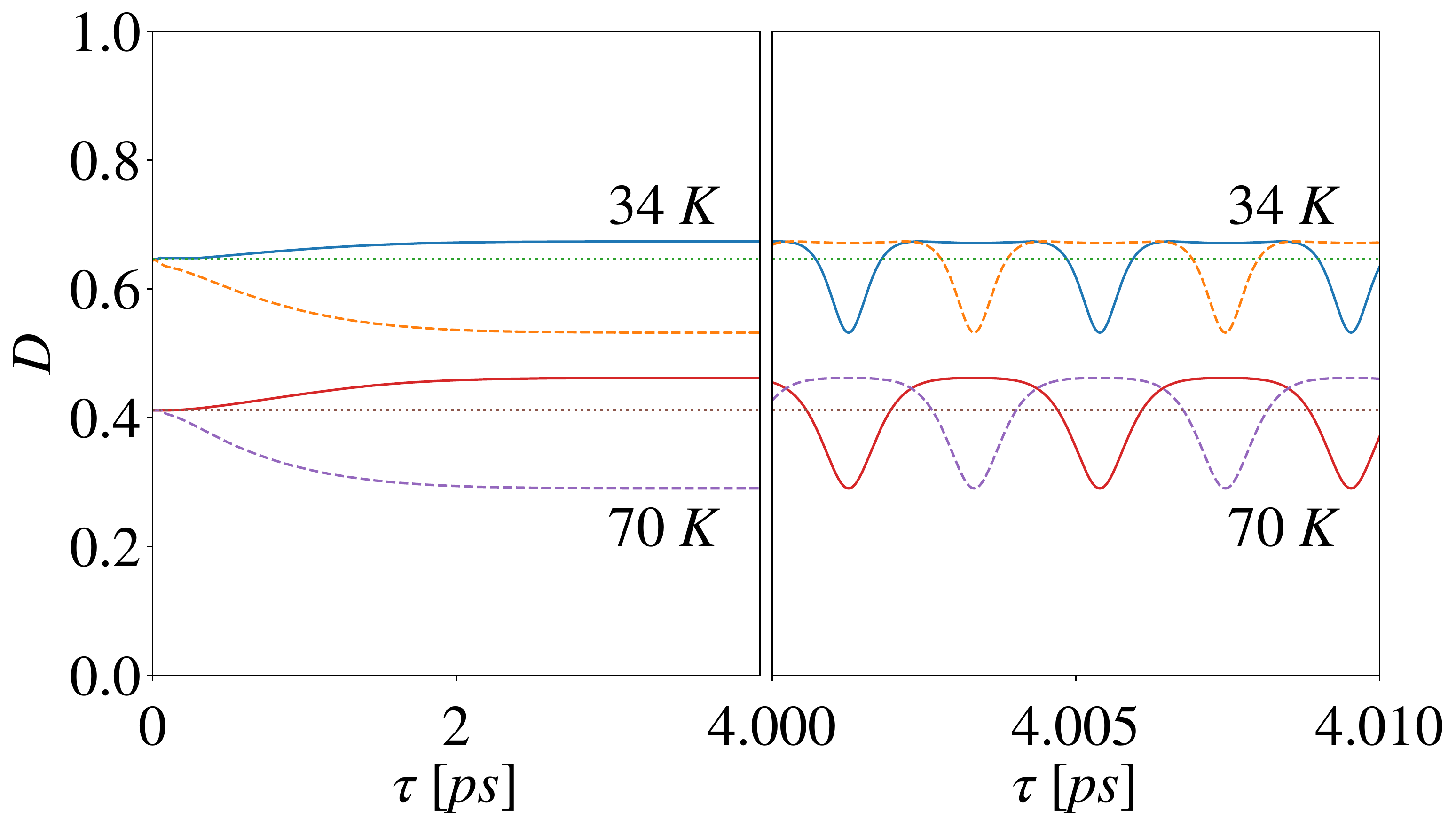}
	\caption{Degree of coherence as a function of the delay time $\tau$ 
		at long $t$ for two temperatures: 34 $K$ and 70 $K$. Left panel shows envelopes
		of the evolution (solid lines - maximum coherence, dashed lines - minimum coherence) and degree of standard decoherence (dotted lines). Right panel shows 
		explicit fast oscillations: solid lines correspond to the $|+\rangle$ measurement outcome, dashed lines - $|-\rangle$ measurement outcome. }
	\label{fig:degree_of_coherence_vs_tau} 
\end{figure}
    
    The Hamiltonian is very similar to the spin-boson model, with the exception 
    of being asymmetric with respect to qubit pointer states, whereas in the spin-boson model
    both the qubit and the interaction terms are proportional to the $\hat{\sigma}_z$
    Pauli operator. The consequence of this is that the exciton-phonon interaction
    has many of the traits characteristic for the spin-boson model which make it 
    so fundamental for the study of decoherence, such as being non-Markovian,
    entangling at finite temperatures \cite{salamon17,rzepkowski21}, and displaying qubit decoherence to a non-zero value
    for super-Ohmic distributions of coupling constants \cite{leggett87,krummheuer02,tuziemski19}. The important difference is
    that, contrary to the spin-boson model, the conditional evolution operators
    of the environment $\hat{w}_i(t)$ do not commute. 
    
    The Hamiltonian under study can be diagonalized exactly using the Weyl operator method
    and the explicit forms of the $\hat{w}_i(t)$ operators which enter the evolution
    operator (\ref{u}) can be found following Refs \cite{roszak06a,rzepkowski21}
    regardless of the form of the coupling constants $f_{\pmb{k}}$.
    These are given by
    \begin{subequations}
    	\label{w01}
    \begin{eqnarray}
	    \label{w0}
	    \hat{w}_0(t)&=&\bigotimes_{\bm{k}}e^{-i\omega_{\bm{k}}\hat{b}_{\bm{k}}^{\dagger}
		\hat{b}_{\bm{k}}t},\\
	    \nonumber
	    \hat{w}_1(t)&=&
	    e^{-i\sum_{\bm{k}}\frac{|f_{\pmb{k}}(t)|^2	
	    	}{\omega^2_{\pmb{k}}}\sin \omega_{\pmb{k}}t}\bigotimes_{\bm{k}}
    	e^{\frac{f_{\pmb{k}}(t)}{\omega_{\pmb{k}}} b_{\pmb{k}}^\dagger 
    		- \frac{f_{\pmb{k}}^*(t)}{\omega_{\pmb{k}}} b_{\pmb{k}}}
    	\hat{w}_0(t),\\
    	\label{w1}
	\end{eqnarray}
	\end{subequations}
	where the time-dependent coupling constants are given by
	\begin{equation}
	f_{\pmb{k}}(t)=f_{\pmb{k}}\left(e^{-i \omega_{\pmb{k}} t}-1\right).
	\end{equation}
	In the operator (\ref{w1}), we have omitted a trivial oscillating phase term, which shifts the excitonic 
	energy difference by a constant term of $\sum_{\bm{k}}\frac{|f_{\pmb{k}}(t)|^2	
	}{\omega^2_{\pmb{k}}}$.
	We include this term directly in said energy difference, so that $\Delta\varepsilon
	=\varepsilon-\sum_{\bm{k}}\frac{|f_{\pmb{k}}(t)|^2	
	}{\omega^2_{\pmb{k}}}$, but it is relevant to note that this shift is very small.

    \subsection{Continuous phonon spectrum\label{4a}}
    
    For an exciton confined in a quantum dot interacting with bulk phonons, the interaction is
    dominated by the deformation potential coupling \cite{krummheuer02,mahan00} and the coupling constants
    are given by 
    \begin{equation}
    f_{\pmb{k}} = (\sigma_e - \sigma_h) \sqrt{\frac{ \pmb{k}}{2 \varrho V_N c}} \int_{-\infty}^{\infty} d^3 \pmb{r} \psi^*(\pmb{r})e^{-i\pmb{k} \cdot \pmb{r}} \psi(\pmb{r}).
    \end{equation}
    Here, $\sigma_{e/h}$ are deformation potential constants for electrons and holes, respectively,
    $\varrho$ is the crystal density, $V_N$ is the phonon normalization volume,
    $c$ is the longitudinal speed of sound (we assume linear dispersion), 
    and $\psi(\pmb{r})$ is the excitonic wave function.
    In the following, we will be using material parameters corresponding to small,
    self-assembled GaAs quantum dots. The wave function is modeled by an anisotropic Gaussian of
    $l_{\perp} = 4$ nm width in the $xy$ plain and $l_z = 1$ nm in the $z$ direction. The difference
    of deformation potential constants is given by $\sigma_e - \sigma_h=9$ eV,
    while $\varrho=5360$ kg/m$^3$ and $c=5100$ m/s. The difference in qubit pointer state
    energies is taken to be $\Delta\varepsilon =1$ eV.
    
    \begin{figure}[t]
    	\includegraphics[width=\linewidth]{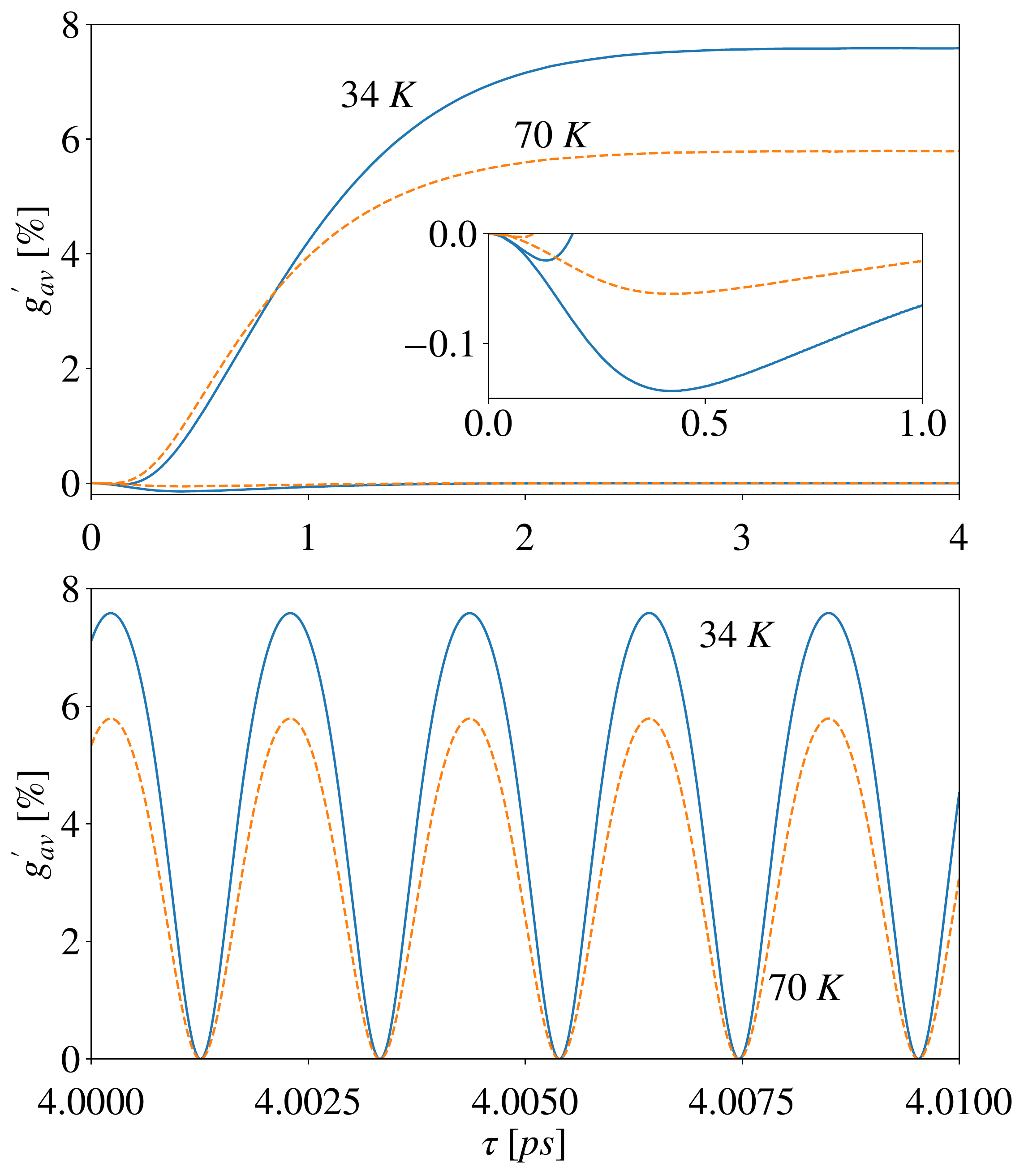}
    	\caption{The average gain of coherence $g_{av}'$ as a function of the delay time at long $t$ for two temperatures: 34 K (blue, solid lines) and 70 K
    		(dashed, orange lines). Top panel presents envelopes of $g_{av}^\prime$ with the inset showing the area, in which $g_{av}^\prime$ reaches global minimum. Bottom panel shows full
    		evolution around $\tau=4$ ps.}
    	\label{fig:average_coherence_gain} 
    \end{figure}
    
    The left panel of Fig.~\ref{fig:degree_of_coherence_vs_tau} shows the maximum and minimum
    values of the degree of coherence at long $t=20$ ps as a function of the delay time $\tau$
    as compared to the standard degree of coherence (dotted, constant line).
    The gain (or loss) stabilizes at a given level after the delay time is longer than a few picoseconds. This threshold value corresponds to the time after initialization when the maximal dephasing is reached. In the right panel of the same figure, the full dynamics are
    shown. The oscillations arise from the interplay of the terms which depend on the 
    coherent evolution of the qubit in the pre-measurement phase (\ref{czescib}).
    For long enought times (here, above $\sim 3.5$ ps) the minima (maxima)
    for the measurement outcome $|+\rangle$ ($|-\rangle$) correspond to the situation
    when $\Delta\varepsilon/=(2j+1)\pi$ (points of minimal gain)
    and the maxima (minima) to $\Delta\varepsilon/=2\pi j$ (point of maximal gain),
    where $j$ is a natural number. A higly relevant point is $\Delta\varepsilon/=(j+1/2)\pi$
    for which the gain in coherence is equal for both measurement outcomes (point of equal gain).
    Interestingly the gain at this point is not much different than that at the point
    of maximal gain and at some temperatures may even by slightly larger (see the $34$ K curves).

    \begin{figure} 
    	\centering
    	\begin{subfigure}{\linewidth}
    		\includegraphics[width=\linewidth]{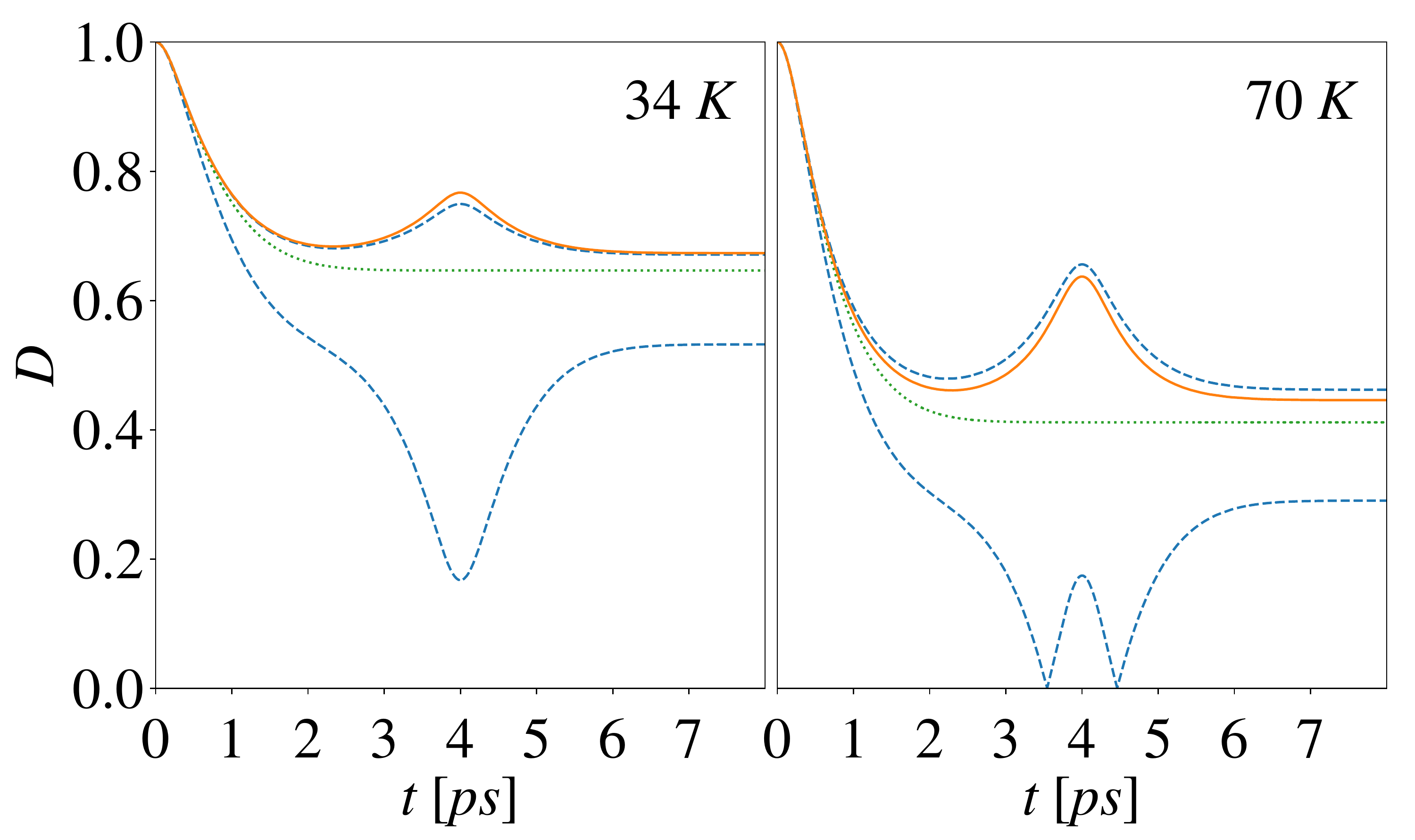}
    		\caption{$\tau_{MIN/MAX}\approx 4$ ps.}
    		\label{fig:degree_of_coherence_vs_t_a} 
    	\end{subfigure}
    	\begin{subfigure}{\linewidth}
    		\includegraphics[width=\linewidth]{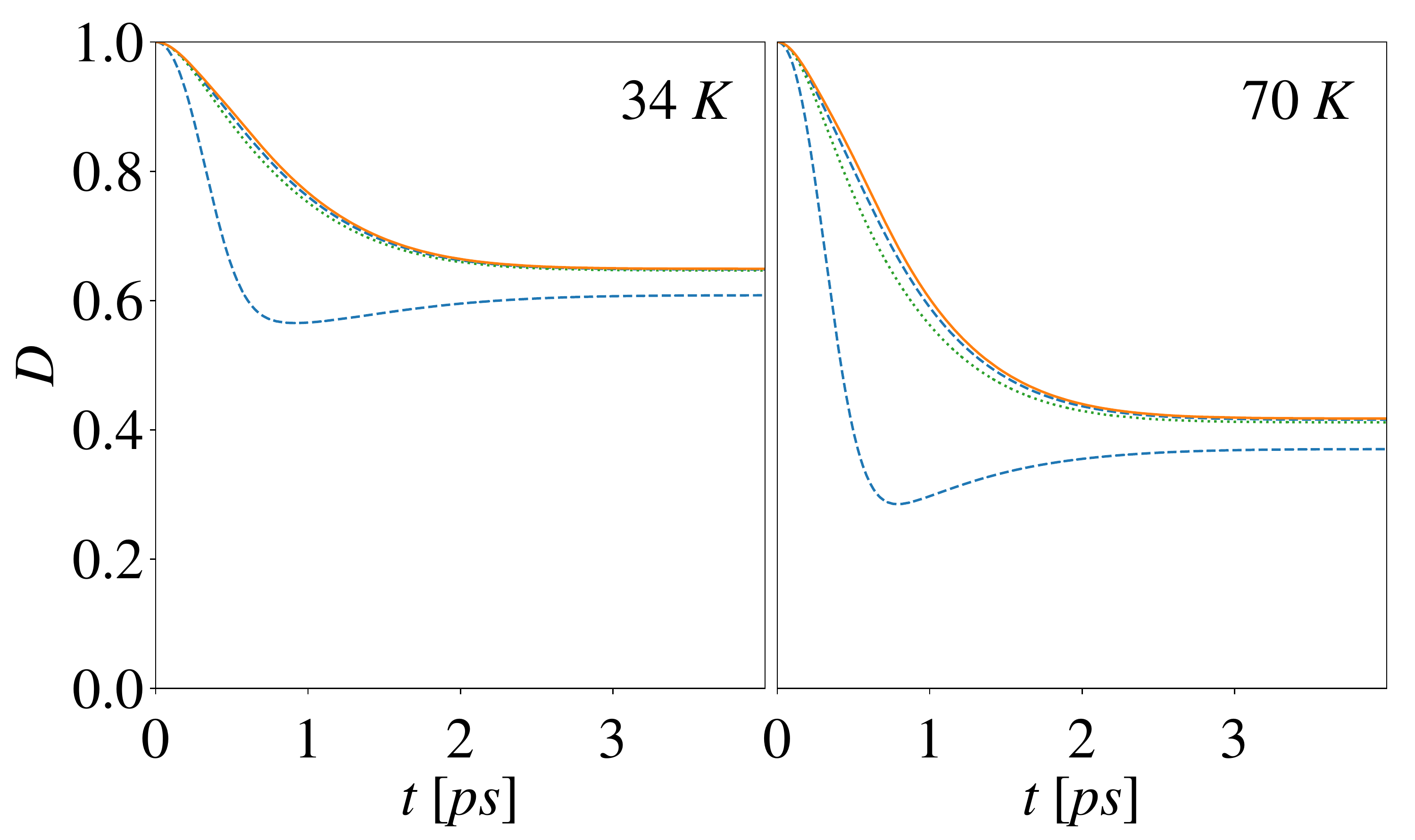}
    		\caption{$\tau_{MIN/MAX}\approx 0.42$ ps.}
    		\label{fig:degree_of_coherence_vs_t_b} 
    	\end{subfigure}
    	\caption{Degree of coherence as a function of time elapsed from the measurement $t$
    		at $T=34$ K (left plots) and $T=70$ K (right plots) for long delay times $\tau\approx 4$
    		ps (top panels) and $\tau\approx 0.42$
    		ps (bottom panels). Dashed lines correspond to points of maximal and minimal gain,
    		while solid lines correspond to points of equal gain. Dotted lines are 
    		standard decoherence curves.}
    	\label{fig:degree_of_coherence_vs_t} 
    \end{figure}

    Fig.~\ref{fig:average_coherence_gain} shows the corresponding average gain of coherence (\ref{average_gain}) normalized by coherence which would be left in the system
    at time $t$ during the stnadard decoherence process, 
    $g_{av}'(\tau,t) = g_{av}(\tau,t) / (1 - D(t))$, yielding a percentage of coherence gained.
    This is plotted at a long decoherence time $t=20$ ns, so all phonon-induced
    processes on the qubit have been finalized. In case of the system under study, the reservoir
    is super-Ohmic, which leads to $1-D(t)$ being fixed at a finite value; only at infinite
    temperature can complete decohrence be obtained \cite{leggett87,an07,hu12,tuziemski19}. 
    
        \begin{figure}[t]
    	\includegraphics[width=\linewidth]{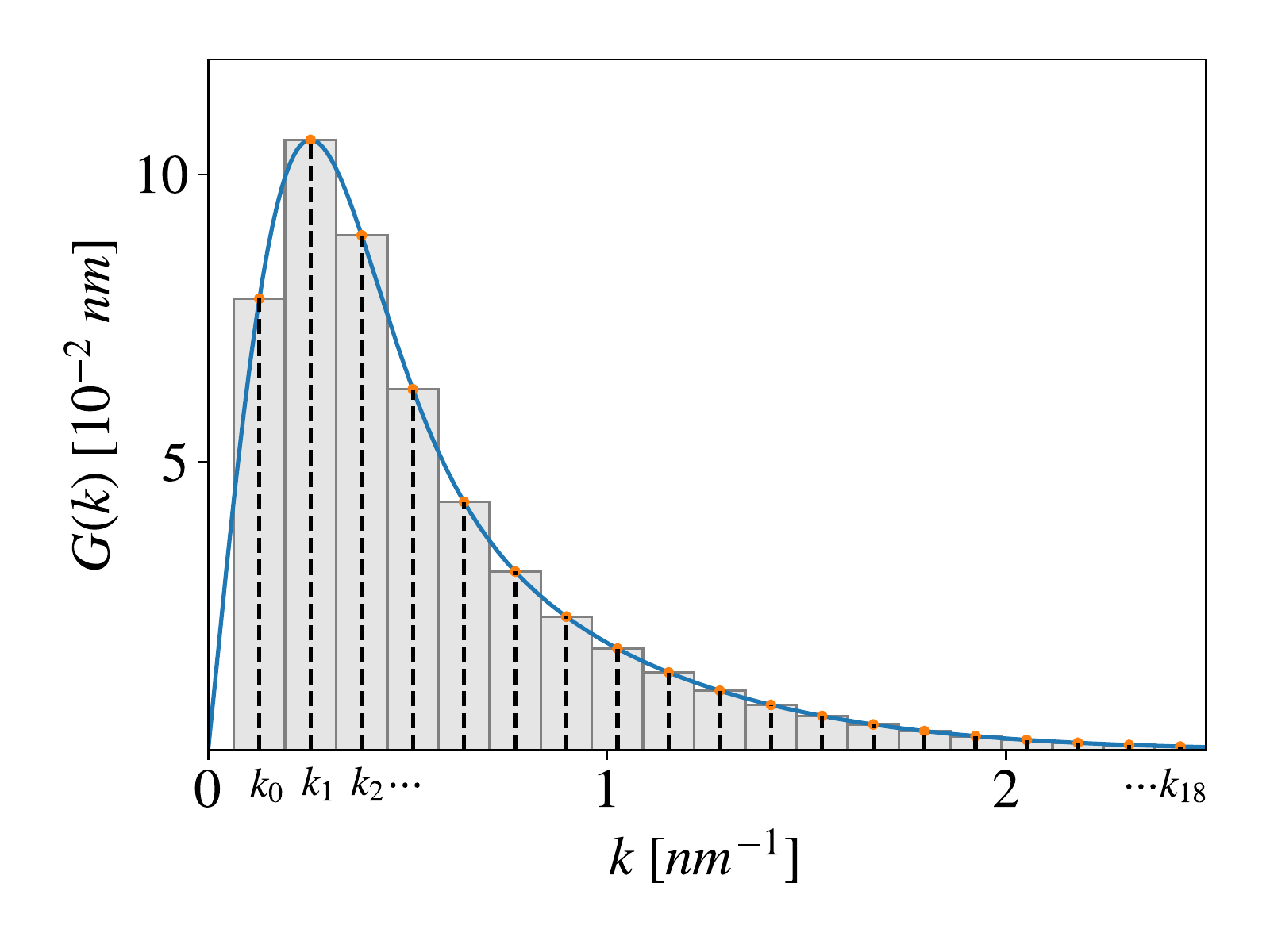}
    	\caption{Function $G(k)$ for continuous phonon spectrum and the 19 phonon modes
    	chosen in the discretization. Bars corresponding to each mode $k_i$
    	have width $\Delta k$.}
    	\label{fig:gk_squared}
    \end{figure}
    
    The top panel of Fig.~\ref{fig:average_coherence_gain} again shows the envelope functions
    of $g_{av}'(\tau,t)$ (the fast oscillations resulting from the large value of $\Delta\varepsilon$ are shown in the bottom panel for large delay times, around $\tau=4$ ps)
    at two temperatures.
    A saturation of the curves can be observed around $3.5$ ps, which corresponds well
    to the time-scales at which charge carrier-phonon processes occur in such systems. 
    At later times the minimum value of the coherence gain is zero, because the non-classical
    phase factors are effectively zero and the expectation values of the environmental
    operators are no longer affected by their lack of commutation.
    The inset shows the short-time evolution when the minimum envelope curves are both below
    zero due to the non-classical phase factors which are relevant at such times.
    Note, that for very short times, both the minimum and maximum curves are below zero,
    so a loss of coherence is seen regardless of the fast oscillations between these 
    curves.
    
    To give the full picture of the phenomenon under study in Fig.~\ref{fig:degree_of_coherence_vs_t} we present the degree of coherence as a function of time that elapsed from the intermediate measurement, $t$. In the upper panels,
    the curves correspond to $\tau\approx 4$ ps, which is well beyond the phonon relaxation times. The actual delay times for the three curves correspond to the 
    points of maximal, minimal, and equal gain, which are closest to $4$ ps.
    Since the envelope functions change slowly in comparison to the oscillatory behavior
    driven by the energy difference between the qubit states, these curves can be 
    freely compared. The time evolutions of the decoherence strongly resemble standard
    evolution (dotted lines) with the exception of a maximum/minimum which is present
    at $t\approx \tau$. This can be understood with the help of eq.~(\ref{bcomut})
    which holds for commuting operators $\hat{w}_i$, where decoherence is cancelled in 
    one of the terms at $t=\tau$, yielding an extreme effect (either positive or negative,
    depending on the measurement outcome) on the degree of coherence. 
    
    The bottom panels show plots corresponding to the top panels, but at short delay times 
    $\tau\approx 0.42$ corresponding to the minimum values in the inset of 
    Fig.~\ref{fig:average_coherence_gain}. The time evolutions are qualitatively different,
    firstly because the curves corresponding to the points of maximum and equal gain
    are almost identical to the standard decoherence decay, and the only distinct 
    evolutions are the ones for the point of minimum gain (which are here always 
    related to a loss).
    Furthermore, only these curves display any distinct behavior at $t=\tau$, which is a minimum,
    much less distinct than in the case of the evolutions corresponding to long delay times
    $\tau$.

	\subsection{Discrete phonon spectrum\label{4b}}    
    
    In this subsection we focus on a situation, in which the quantum dot interacts with a finite number of phonon modes, corresponding to the situation when the phonons are confined to a small
    structure and cannot be treated as those propagating within a bulk crystal \cite{mahan00}. 
    To obtain qubit evolutions of a similar magnitude and occuring on similar time-scales
    as in the previous section, we will not model a specific medium for the phonons,
    but instead artificially ``quantize'' the phonon spectrum in terms of the phonon
    wave number, $k=|\pmb{k}|$. 
    
    This procedure is illustrated in Fig.~\ref{fig:gk_squared},
    where the continuous line shows the $k$-dependence of $G(k)$, which is defined by
    the equation
    $H=\sum_{\pmb{k}} | \frac{f_{\pmb{k}}}{ \omega_{\pmb{k}}}|^2=\int_0^{\infty}G(k)$
    using the parameters from the previous subsection (\ref{4a})
    and is explicitly given by
	\begin{eqnarray} \label{sum_to_integral_transition}
		G(k) &=& \frac{NV}{(2 \pi)^2} \frac{(\sigma_e - \sigma_h)^2}{2 \rho  c^3} k\int_{0}^{\pi} \,d\Theta  sin \Theta \nonumber \\
		&& \times \exp[-\frac{1}{2}(l_z^2 k^2 \cos^2 \Theta + l_{\perp}^2 k^2 \sin^2 \Theta)].
	\end{eqnarray} 
	Here the integration over the azimuthal angle has been performed and only
	the integration over the polar angle remains. 
	We then choose $19$ equally spaced values of the wave number $k_i$ with $i=0,1,\dots, 18$,
	where $k_1$ corresponds to the maximum of $G(k)$ and $\Delta k=k_{i+1}-k_{i}=2/3k_1=
	0.1282$ nm$^{-1}$,
	which means that the $k=0$ mode is taken into account with $G(k=0)=0$.
	We can now find $H_i=| \frac{f_{k_i}}{ \omega_{k_i}}|^2=G(k_i)\Delta k$   
	and the corresponding values of the coupling constants
	$f_{k_i}= \omega_{k_i}\sqrt{H_i}$ which enter eqs (\ref{w01}). 
	This allows for all of the relevant quantities for the scheme under study to be
	found after replacing the integration over the wave vector $\pmb{k}$
	with a summation over wave numbers $k_i$. In fact, only the particular values of $H_i$
	and the corresponding energies $\omega_{k_i}$, with $\omega_{k_i}=ck_i$
	under linear dispersion,
	are necessary to find the evolution of the degree of coherence both for standard decoherence and when the decoherence reduction
	scheme is used. Note that $H/\sum_iH_i=1.0176$ so that the bars in fig.~\ref{fig:gk_squared}
	represent the function very well. 
    
    \begin{figure}[t]
    	\includegraphics[width=\linewidth]{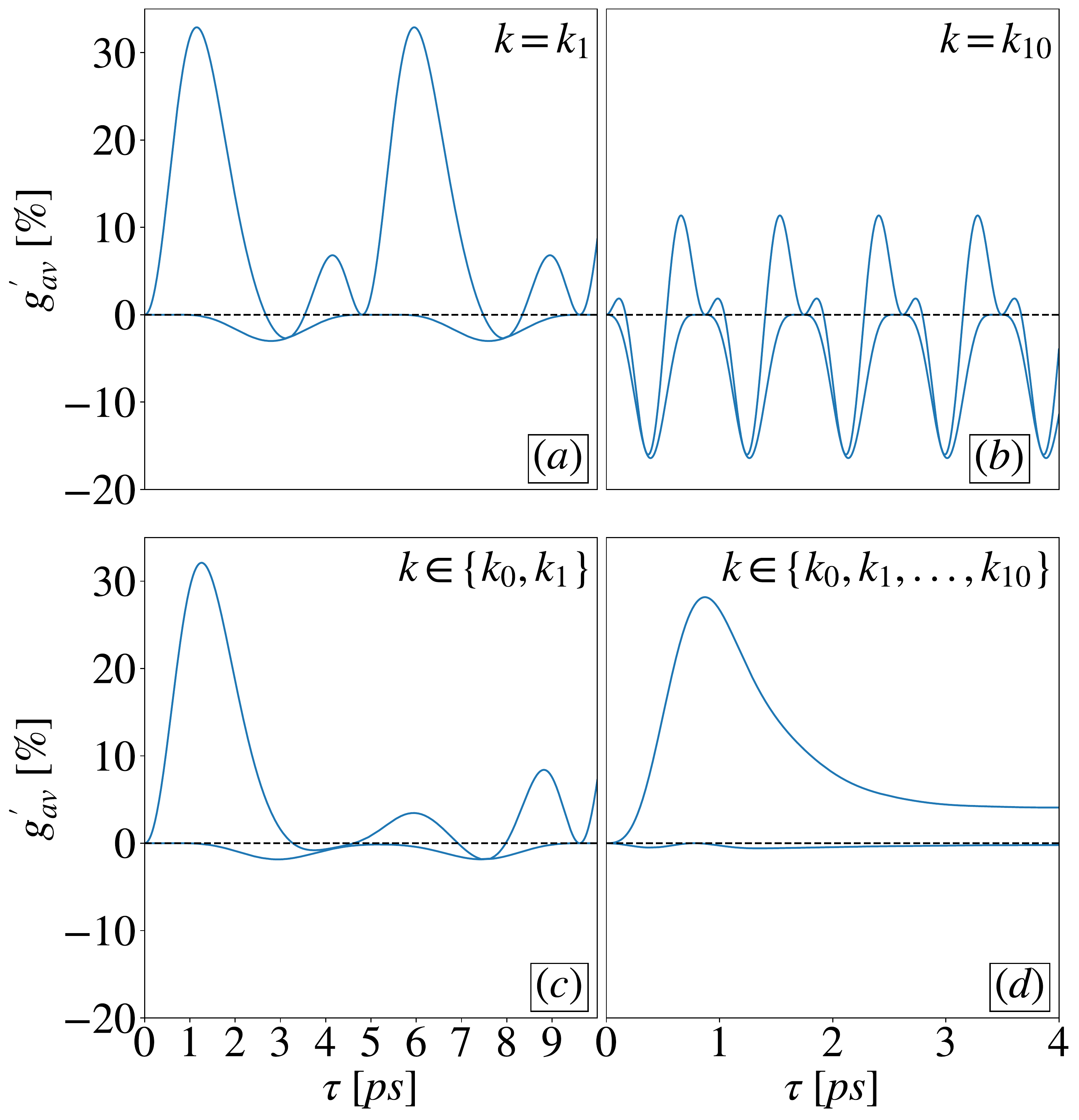}
    	\caption{Envelope functions of the average gain of coherence: dependence on
    		delay time $\tau$
    		at long $t=20$ ps for selected discrete sets of phonon wave numbers $k$. 
    		Environment with a single phonon mode:
    		a) $k=k_1$ and b) $k=k_{10}$. More environmental modes: c) two phonon modes, $k_0$ and $k_1$,
    		c) eleven phonon modes $k_i$, $i=0,\dots,10$.}
    	\label{fig:average_coherence_gain_discrete_ks}
    \end{figure}

	Fig.~\ref{fig:average_coherence_gain_discrete_ks} shows the $\tau$-dependence
	of the average gain of coherence for $t=20$ ps analogous to Fig.~\ref{fig:average_coherence_gain}, but for a choice of wave numbers out of the 19
	specified in Fig.~\ref{fig:gk_squared} and described above. The coupling strengths are always adjusted, so that $\sum_iH_i=H$. When only a few phonon modes are taken into 
	account (a,b,c), the evolution is highly 
	periodic and the minima of the coherence gain display large
	negative values (especially in comparison with all that is observed in the case
	of the continuous spectrum). The recurring negativity is due to the fact that the
	effective cancellation of the non-commutativity of the operators (\ref{w01})
	is connected with the interplay of many phonon modes when the spectrum approaches
	continuum. Note that this effect is obtained already for 10 phonon modes,
	as seen in fig.~\ref{fig:average_coherence_gain_discrete_ks} (d). For smaller systems 
	such cancellation cannot occur and the non-classical phases which lead to negative
	values of the coherence gain are much larger. 
	
	The top panels contain evolutions resulting from an interaction with only a single
	phonon mode, $k=k_1$ (a) and $k=k_{10}$ (b). Due to the adjustment of the coupling strength,
	they both interact similarly with the qubit
	and one can observe a change of periodicity of the evolution which is the result of the
	different energies of the two phonon modes, $\omega_{k_i}$.
	Qualitatively, the evolution is very similar, displaying two maxima of different magnitude
	and a minimum in each period. The differences are strongly visible in the gain
	and loss which can be achieved in the two cases. The reason for these differences lies
	in the arbitrary choice of the constant time $t$. Since both evolutions are periodic,
	but with a different periodicity, the interplays of times $t$ and $\tau$ are different
	in the two cases, yielding one evolution which is more favorable and has a high maximum
	gain, and one which is much worse in terms of coherence gain and a lot of the time 
	leads to loss. 
	
	Adding a second phonon mode as in fig.~\ref{fig:average_coherence_gain_discrete_ks} (c)
	leads to an expectedly more complicated evolution with three maxima an two minima
	in a single period, whereas the initial evolution is still dominated by the $k_1$ mode. With each additional mode this trend continues simultaneously extending the period of the 
	evolution. Already at 10 modes (d), the losses of coherence become negligible
	in comparison to the situation when only a few modes are present and the tendency
	toward decoherence stabilizing at a finite value is demonstrated. This value is similar
	as in in the continuous spectrum case, Fig.~\ref{fig:average_coherence_gain}, but a strong
	maximum characteristic for few-mode evolutions is still present. At 19 phonon modes,
	the evolution is of the same character as in fig.~\ref{fig:average_coherence_gain_discrete_ks} (d).
	
	\section{Conclusion \label{sec5}}
	
	We have generalized the scheme for the reduction of decoherence introduced for a charge qubit
	interacting with a phononic bath \cite{roszak15b} to the case
	of all qubit-environment interactions that lead to pure decoherence of the qubit.
	In this framework, we have shown that the scheme will always result in gain of coherence
	on average (over two outcomes of the qubit measurement which is the core of
	the scheme) if a mixture of environmental states $1/2[\hat{R}_{00}(\tau)+\hat{R}_{11}(\tau)]$
	leads to the same decoherence curves as the initial environmental state $\hat{R}(0)$.
	The density matrices $\hat{R}_{ii}(\tau)$ correspond to environmental states
	which would be obtained from the initial state $\hat{R}(0)$ if the environment
	interacted with the qubit in state $|i\rangle$ for time $\tau$.
	
	Such a situation is most commonly encountered when the operators that govern
	the evolution of the environment conditional on the pointer state of the qubit
	$|i\rangle$, commute. This is usually the outcome of commutation between 
	different parts of the Hamiltonian, responsible for the distinguishability of 
	qubit states by the environment. Another possibility is when the conditional environmental
	evolution operators do not commute with each other, but do commute with the 
	initial state of the environment. For mixed states, this does not exclude decoherence,
	even though it excludes the formation of qubit-environment entanglement \cite{roszak15}.
	An example here is the infinite-temperature density matrix, which commutes with
	any evolution operator, and typically yields the strongest decoherence. 
	
	We observe this type of behavior in our example of a charge qubit coupled
	to an environment of phonons only for a continuous phonon spectrum and at long times,
	both the time of measurement and the time elapsed afterwards. At finite temperatures,
	this system does not fulfill any of the commutation relations specified in the
	previous paragraph, but nevertheless displays only gain of coherence and no loss.
	This is because at long times the non-trivial phase oscillations, which are the
	outcome of the lack of commutations, cancel out due to large-system effects.
	Hence, there are more situations where loss of coherence is unlikely, which
	are typically connected with more classical properties of the environment.
	
	Situations when loss of coherence is observed are of more fundamental interest
	since they are only possible when the conditional evolution operators
	of the environment, and consequently observables which correspond to 
	different parts of the Hamiltonian, do not commute.
	Hence, negative average gain of coherence is a witness of quantumness of the
	Hamiltonian and of the evolution of the environment, even though
	it does not have to be accompanied by qubit-environment entanglement at time $\tau$
	[when $\hat{R}_{00}(\tau)=\hat{R}_{11}(\tau)<\hat{R}(0)$].
	
	We observe very small losses of coherence at small times for the continuous spectrum
	example, but it is very common for a small and discrete number of phonon modes. 
	Although for examples chosen in such a way that the same mode dominates the spectrum
	in the continuous and discrete case, observed losses of coherence are small
	and gain is more probable, we were able to find situations when
	the gain is predominantly negative. 
	
    Overall, we have shown that 
    the scheme for pure decoherence control is of much wider usage than
    only for phonon-induced decoherence, as it is able to slow or reduce decoherence
    for a large class of interactions that lead to pure decoherence. In other cases
    (when loss of coherence on average is possible at certain 
    measurement times $\tau$), it can still be used for decoherence control
    if time of measurement can be controlled with some reliability. 

    \section*{Acknowledgement}
    K.~R.~ acknowledges support from project 20-16577S of the Czech Science Foundation.
    B.~R.~ acknowledges support by the European Union under the European
    Social Fund.
    
    \section*{Code avaliability}
    All code used to generate plots in Sec.~\ref{sec4} is available at
    \url{https://github.com/brzepkowski/coherence_gain}.

\end{document}